\documentclass[aps,prd,twocolumn,superscriptaddress]{revtex4-2}

\usepackage{amsmath}

\usepackage{amsfonts}
\usepackage{amssymb}
\usepackage{mathtools}
\usepackage{dsfont}
\usepackage[mathscr]{eucal}
\usepackage{braket}
\usepackage{hyperref}
\hypersetup{colorlinks,%
citecolor=red,%
filecolor=black,%
linkcolor=blue,%
urlcolor=blue}

\newcommand{\rme}{\ensuremath{\mathrm{e}}}
\newcommand{\rmi}{\ensuremath{\mathrm{i}}}

\usepackage[usenames,dvipsnames]{xcolor}
\usepackage{soul}

\DeclareMathOperator{\sign}{sign}

\begin{document}

\title{Transformation of Spin in Quantum Reference Frames}

\author{Marion Mikusch}
\email[]{marion.mikusch@univie.ac.at}
\affiliation{Faculty of Physics, University of Vienna, Boltzmanngasse 5, A-1090 Vienna, Austria}
\affiliation{Vienna Center for Quantum Science and Technology (VCQ)}

\author{Luis C.\ Barbado}
\email[]{luis.cortes.barbado@univie.ac.at}
\author{\v{C}aslav Brukner}
\email[]{caslav.brukner@univie.ac.at}
\affiliation{Faculty of Physics, University of Vienna, Boltzmanngasse 5, A-1090 Vienna, Austria}
\affiliation{Vienna Center for Quantum Science and Technology (VCQ)}
\affiliation{Institute for Quantum Optics and Quantum Information Vienna (IQOQI)}

\date{\today}

\begin{abstract}
In physical experiments, reference frames are standardly modelled through a specific choice of coordinates used to describe the physical systems, but they themselves are not considered as such. However, any reference frame is a physical system that ultimately behaves according to quantum mechanics. We develop a framework for rotational (i.e.\ spin) quantum reference frames, with respect to which quantum systems with spin degrees of freedom are described. We give an explicit model for such frames as systems composed of three spin coherent states of angular momentum~$j$ and introduce the transformations between them by upgrading the Euler angles occurring in classical~$\textrm{SO}(3)$ spin transformations to quantum mechanical operators acting on the states of the reference frames. To ensure that an arbitrary rotation can be applied on the spin we take the limit of infinitely large~$j$, in which case the angle operator possesses a continuous spectrum. We prove that rotationally invariant Hamiltonians (such as that of the Heisenberg model) are invariant under a larger group of quantum reference frame transformations. Our result is the first development of the quantum reference frame formalism for a non-Abelian group.
\end{abstract}

\maketitle

\section{\label{sec:level1}Introduction}

The description of physical systems relies heavily on the choice of the reference frame used. Although practically it is always a physical system that constitutes a reference frame, within the theory they are commonly described as abstract entities that for themselves do not carry any degrees of freedom. A well-known example is the laboratory frame. It is modelled as an idealized mathematical coordinate system of infinite extent that does not possess a physical manifestation and therefore itself is not requested to obey the laws of physics.

However, in any experimental observation a physical system is used as a reference frame. As such, it inherently behaves according to the laws of quantum mechanics. Due to this fact, there have been attempts to quantize the classical notion of reference frames and the concept of so-called quantum reference frames (QRFs) has been extensively studied in the literature.

In order to circumvent superselection rules and develop quantum information tasks when parties have bounded reference frames or even lack them, seminal works proposed reference frames to be treated within the formalism of quantum mechanics~\cite{aharonov1, aharonov3, kitaev1, bartlett4, bartlett1, poulin1, spekkens1, bartlett2, skotin1, bartlett3, smith1}. Methods were developed to encode quantum information in relational degrees of freedom~\cite{bartlett2, poulin2, poulin3, loveridge1, pienaar1, loveridge2, loveridge3}, which upon measurement lead to a degradation of QRFs ~\cite{poulin2, poulin3}. Other authors discuss a possible application of QRFs in quantum gravity~\cite{rovelli1}. Angelo \textit{et.~al.}~\cite{angelo1, angelo2, angelo3} revised the initial work on QRFs~\cite{aharonov1, aharonov3} and examined the quantum description from the perspective of a single quantum particle.

A recent work by Giacomini, Castro-Ruiz and Brukner~\cite{brukner1} outlines a formalism to treat reference systems with quantum degrees of freedom and provides an approach to the transformation of spatial and momentum variables in QRFs that is genuinely relational by construction. The key methodological step is that the parameter of the transformation (e.g. displacement of the translation, velocity of the boost) is replaced by the quantum operator acting on the QRF, so that the transformation generalizes the usual coordinate transformations to a ``superposition of coordinate transformations''. Some of these QRF transformations have subsequently been rederived starting from a gravity inspired symmetry principle in a perspective neutral model~\cite{hohn1} and extended to three-dimensional problems with translational and rotational invariance~\cite{hohn2}.

In an independent approach, Pienaar has provided the first attempt to consider QRFs for spin and derived a symmetry transformation between spin systems~\cite{pienaar1}. Subsequently, the operational meaning of spin in relativistic QRFs has been outlined in~\cite{brukner2}. This work introduces generalizations of the Lorentz boosts between QRF attached to massive relativistic particles. Recently, temporal versions of QRFs, i.e. QRFs associated to quantum clocks, have been introduced~\cite{smith2, castro1, hoehn2} and, using QRFs attached to freely falling quantum systems, Einstein's Equivalence Principle has been generalized to superpositions of gravitational fields~\cite{giacomini1}.

De la Hamette and Galley recently found a group theoretical approach to transformations between QRFs by associating a certain symmetry group to a QRF~\cite{hamette1}. They introduced a generic operator that generalizes a change of QRF for arbitrary symmetry groups. The group theoretical approach to QRF has been extended in~\cite{krumm1, ballesteros1}. Nevertheless, an explicit treatment of reference systems with rotational or spin degrees of freedom has hitherto not been developed.

In this work, we propose the notion of a QRF for quantum spin degrees of freedom and establish a formalism for transformations between them. More specifically, we define spin reference frames as systems composed of three orthogonal spin coherent states (SCS) with infinitely large spin. We `quantize' rotational reference frames by treating the Euler angles entering the classical~$\textrm{SO}(3)$ spin transformations as quantum mechanical operators. This extends the group of transformations between QRFs from rotations to ``superpositions of rotations''. A special case of such a transformation is one between two QRFs that are in a Schr\"odinger-cat like state with respect to each other. We find the dynamical law that is invariant under the extended group of transformations. Finally, we observe that superposition and entanglement are reference-frame dependent notions, in agreement with the results of~\cite{brukner1} for spatial position and momentum degrees of freedom.

We consider our QRF to have unlimited resources for measuring orientations, hence the angular momentum quantum number~$j$ of the SCS constituting a QRF is infinitely large. Therefore, not every spin quantum system is considered suitable to act as a QRF for rotations in this work. Rather, only the systems consisting of three SCS with infinite spins aligned in three orthogonal directions in space are considered as such. This does not mean, however, that we take the classical limit, since we can still consider superpositions of QRFs in this limit (similarly to Schr\"odinger-cat states). The limit $j \rightarrow \infty$ enables us to perform arbitrary conditional~$\textrm{SO}(3)$ rotations on the physical system being described within the different reference frames. Note that the assumption of unlimited resources was tacit also in the framework for QRF transformation of spatial position and momentum variables developed by Giacomini \textit{et.~al.}~\cite{brukner1}. Notice also that we use SCS with infinitely large spin as a resource for measuring directionality that we find particularly convenient in the framework of spin transformations. However, any other states which could act as resources of directionality may in principle be equally valid for this purpose. In particular, states with well-defined three momentum of the (rigged) Hilbert space for the spatial degrees of freedom of a particle could also be used, at least formally.

While QRFs transformations are mathematically well-defined and self-consistent, one may wonder what operational meaning can be attached to such transformations. What does it mean physically to ``jump'' to a quantum reference frame and view external systems from its perspective? Naively, this seems impossible, since we are inherently stuck in a classical ``laboratory frame''. There are two different ways to understand transformations between (quantum) reference frames. An \textit{active} transformation is one in which the state of a physical system is actually changed. A \textit{passive} transformation, on the other hand, refers to different state assignments in different reference frames without changing the system.

For an active QRFs transformation, one would need to prepare a macroscopic (reference frame) system in a quantum state with respect to the laboratory and then consider measurements on other systems relative to that system. For example, it has been suggested that a vibrating wire can serve as a quantum inertial reference frame \cite{PhysRevA.92.042104}. This reference frame can be extended by an atom placed on it, which can "measure" external systems (including parts of laboratory devices such as mirrors) by absorption and emission of photons. It is clear that it is experimentally very difficult to achieve such macroscopic superpositions, although the limits of preparing larger and heavier systems in quantum states are constantly being pushed \cite{PhysRevLett.127.023601}.

In a passive transformation, we can always take mentally the point of view of a reference frame and analyze other systems mathematically as if these systems were actually observed from that reference frame. The method has become standard in physics. Evidence for the time dilation predicted by special relativity, for example, has been provided by experiments measuring the lifetime of muons traveling at about 0.99c, which are caused by the collision of cosmic rays with the upper atmosphere. If there were no time dilation, then the muons would have to decay still in the upper regions of the atmosphere, but as a result of time dilation they are present in significant quantities even at much lower altitudes. To quantify this effect, we can mentally go to the muons' rest frame, for which we know their lifetime, and then transform it relativistically to obtain the lifetime in the Earth system. Of course, no observation was actually made from the muons' rest system. The same method was recently used to define the spin operator of a massive particle that moves in a superposition of relativistic velocities in a laboratory reference frame. Using a passive transformation, we mentally transform to the rest frame of the particle, define the spin operator there and transform it back to the laboratory frame. In a similar vain in \cite{krumm1} the authors propose an operational interpretation of QRF choice as equivalent to ``aligning one’s description of the physics with respect to some choice of internal
quantum subsystem''.

In the present work, QRFs are represented by semi-classical states of infinitely large spins (spin coherence states) and their superpositions. The experimental realization of such superpositions is currently unattainable, but could be achieved in the near future for moderate spin lengths. We refer to experimental results in the field of nuclear quadrupole resonance \cite{Auccaise_Estrada_2013,2018QuIP...17..177T}. Finally, we mention that in the limiting case of large spins, the SCSs are formally equivalent to the optical coherent states \cite{PhysRevA.67.042113}. The development of superposition and entanglement of optical coherent states have been studied in a much more advanced stage, both theoretically and experimentally (see, e.g., \cite{Sanders_2012,Israel:19}).

The article is organized as follows: In Section~\ref{sec:level2} we present the main results of the work. We first introduce the SCS and the transformation between classical reference frames for spins, and then combine these two ingredients to build the transformations between QRFs. We complete the Section by providing some illustrative examples of transformations between QRFs for spin. In Section~\ref{sec:invar} we prove the invariance of the rotationally invariant Hamiltonians under the larger set of QRF transformations. Finally, in Section~\ref{sec:level4} we conclude with a physical discussion of the results and some possible generalizations.

\section{\label{sec:level2}Results}

Let us focus on the following physical situation: A and C are two arbitrary QRFs for spin, which we will precisely define later, and B is a spin quantum system. As a starting point we imagine ourselves 'sitting' in the frame C and assume that from this perspective we describe the physical properties of both A and B. Then, we change to the viewpoint of A. The key questions we address in this paper are: How spin degrees of freedom are encoded in the physical structure of the QRFs A and C, and how such a change of perspective between them induces a transformation of the description of B and the remaining QRF.
\begin{figure}
\includegraphics[width=0.7\linewidth]{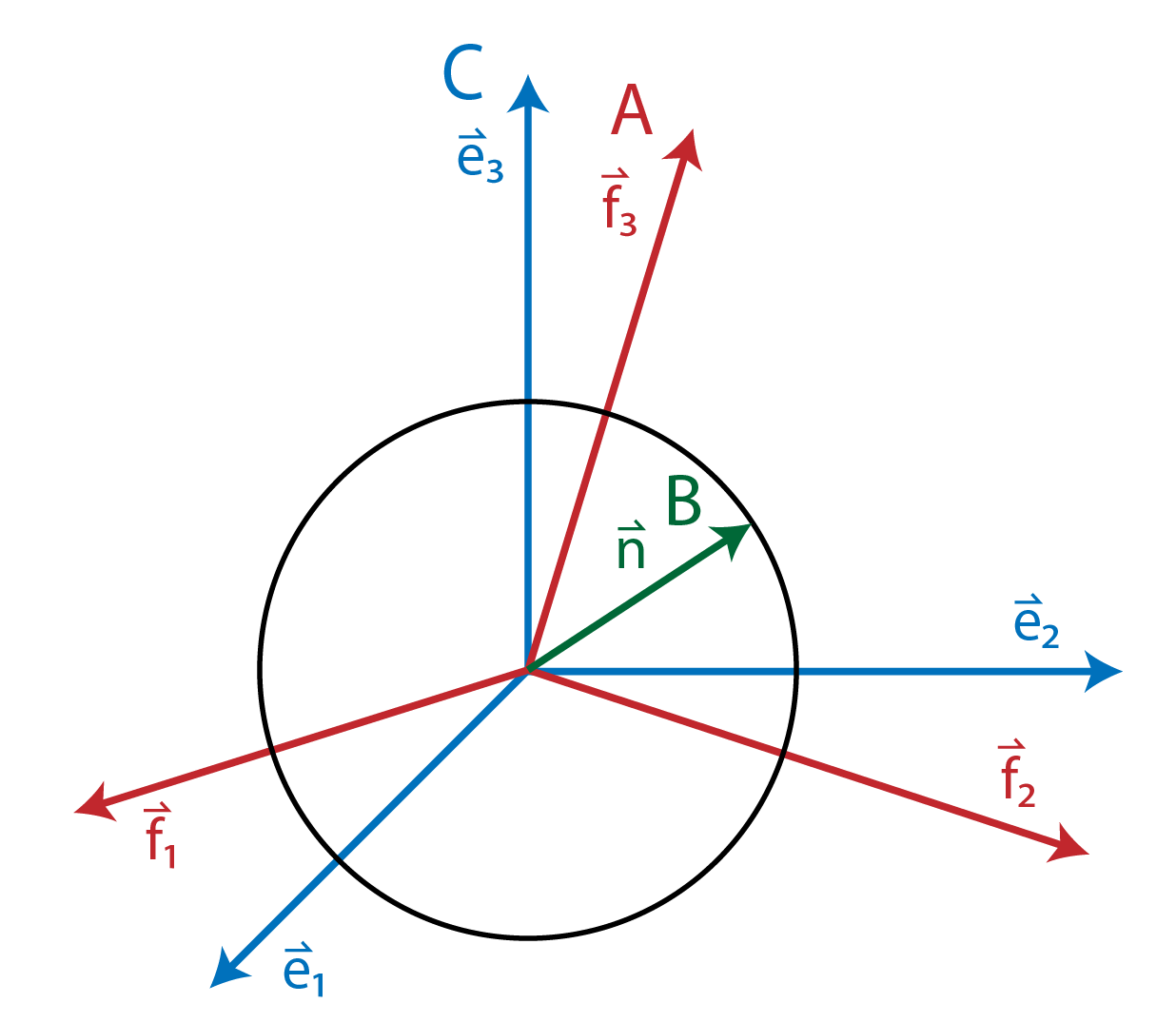}
\caption{Image of two QRFs A and C and a quantum spin system B. A and C are illustrated as two classical coordinate systems but should be thought of as quantum laboratories equipped with their own devices allowing for arbitrarily precise measures of orientations. Here, the quantum system B is depicted as a Bloch vector $\vec{n}$. \label{fig1}}
\end{figure}

In \textit{Figure~\ref{fig1}} the usual `classical reference frame' version of this situation is depicted. Under a classical change of reference frame a physical system exhibits a coordinate transformation in which the new coordinates are expressed in terms of the old ones. In the case of rotational degrees of freedom, this transformation constitutes a representation of the underlying symmetry group~$\textrm{SO}(3)$ and is characterized by three successive rotations given by Euler angles. One could also consider transformations between frames that do not share the same chirality, in which case the transformations correspond to the symmetry group~$\textrm{O}(3)$ and include both rotations and reflections. For simplicity, from here on we consider only transformations between reference frames that share the same chirality (both in the classical and the quantum scenarios). We leave the description of transformations between reference frames with different chirality for Appendix \ref{euler}.

In the framework of quantum mechanics, a single spin rotation is represented by a unitary operation $\hat{U}=\rme^{\rmi\phi\vec{n}\cdot\hat{\vec{J}}}$, where~$\vec{n}$ and~$\phi$ are the axis and angle of rotation and $\hat{\vec{J}}=(\hat{J}_x,\hat{J}_y,\hat{J}_z)$ denotes the angular momentum operator. The transformation of a quantum system from one classical reference frame to another would consist of such rotation, plus a conditional reflection.

In order to establish a formalism for spin QRFs, we next introduce the basic concepts of our approach. We define a QRF for spin and adopt the view that the QRF itself is not a dynamical variable with respect to its own description (to avoid `self-reference problem')~\cite{dalla1, breuer1}. Thus, if we imagine ourselves 'sitting' in the frame C, we only perceive the composition of the QRF A and the quantum system B as the `external world' (see \textit{Figure \ref{fig1}}). A and C themselves will be composite spin quantum systems. As we already mentioned, a single QRF will be represented by three SCS in the limit of large spin lengths, which are oriented along three mutually orthogonal directions. The three SCS are the counterparts of the three directional degrees of freedom of an usual (classical) reference frame.

\subsection{\label{subsec:level1}Construction of QRFs for spin}

Let us first proceed to the explicit construction of a QRF for spin by using three mutually orthogonal SCS. A coherent state of a spin-$j$-particle (SCS) along the direction $\vec{n}$, which we denote $\ket{j,m=j}_{\vec{n}}$, is defined as the state with the highest quantum number $m=j$ along such direction $\vec{n}$; i.e. $\hat{J}_{\vec{n}}\ket{j,m=j}_{\vec{n}} = \hbar j\ket{j,m=j}_{\vec{n}}$, where $\hat{J}_{\vec{n}}$ is the angular momentum operator along the direction~$\vec{n}$. From here on, we will denote the SCS shortly as $\ket{\vec{n}}$. The state is usually pictured as a vector living on a Bloch sphere with radius $j$. Mathematically, SCS are associated with unitary irreducible representations of the rotation group~$\textrm{SO}(3)$, and can be represented by
\begin{widetext}
\begin{equation}
\ket{\vec{n}(\theta,\phi)}=\sum_{m=- j}^j\sqrt{\binom{2j}{j+m}}\left(\cos{\frac{\theta}{2}}\right)^{j+m}\left(\sin{\frac{\theta}{2}}\right)^{j-m}\rme^{\rmi(j-m)\phi}\ket{j,m}, \label{spincoherent}
\end{equation}
\end{widetext}
where $\phi$ is the azimuthal angle and $\theta$ the polar angle of the vector $\vec{n}$. An in-depth discussion of SCS and their properties can be found in~\cite{radcliffe1,gazeau1}.
\begin{figure}
\includegraphics[width=0.7\linewidth]{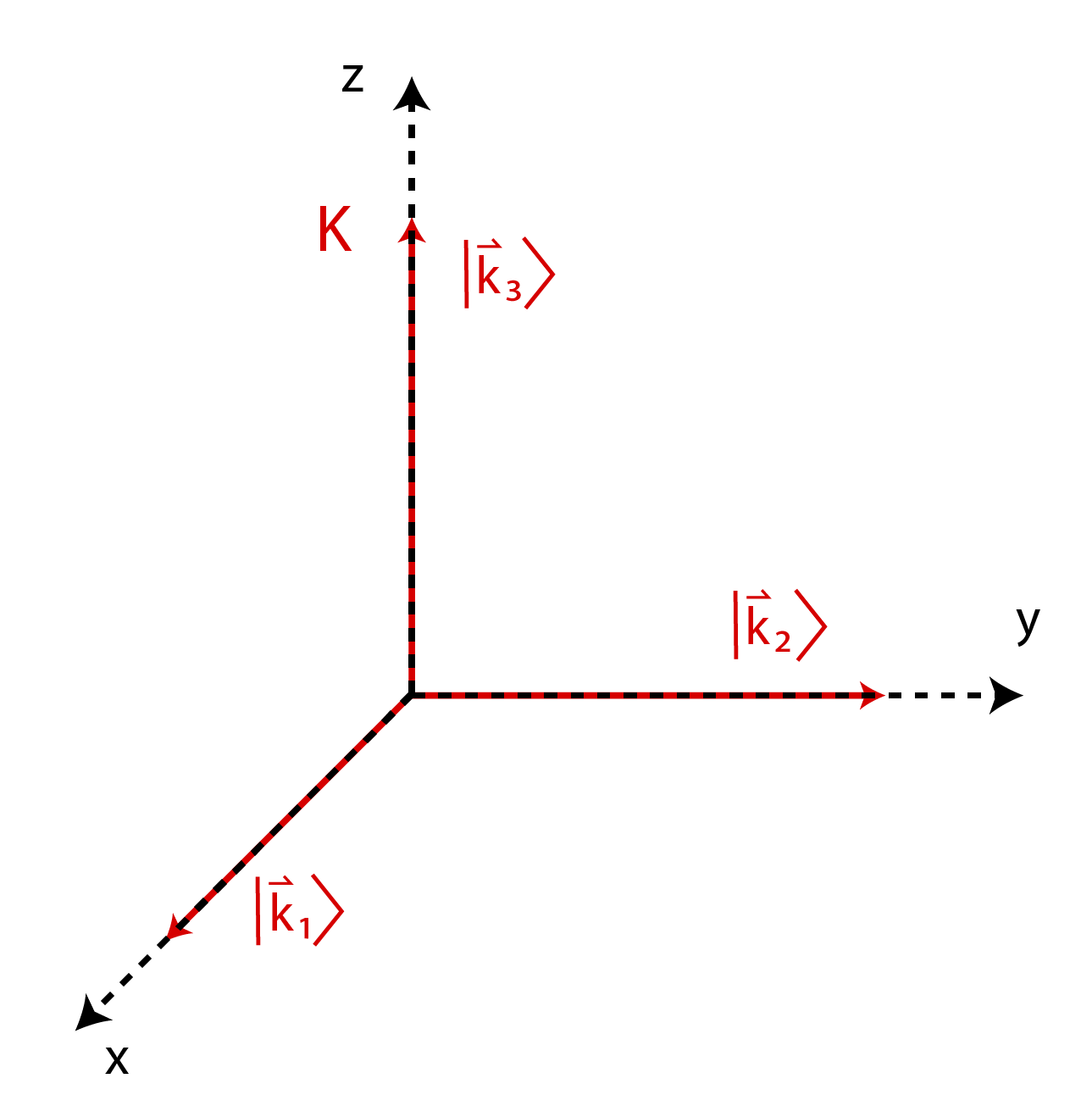}
\caption{Illustration of a QRF for spin with well-defined orientation: The QRF~K is modeled as a composition of three macroscopic spin superposition states (SCS) $\ket{\vec{k}_i}$, $i=1,2,3$, that, in the limit of large quantum numbers $j \rightarrow \infty$, coincide with three mutually orthogonal directions, which would correspond to the coordinate axes $x$, $y$ and $z$ of the reference frame.\label{fig2}}
\end{figure}

We propose that the state of a QRF K for spin \emph{with well-defined orientation} with respect to another QRF consists of three SCS $\ket{\vec{k}_1}$, $\ket{\vec{k}_2}$ and $\ket{\vec{k}_3}$ oriented along three orthogonal directions $\vec{k}_1$, $\vec{k}_2$ and $\vec{k}_3$, respectively (see \textit{Figure \ref{fig2}}), in the limit of $j \to \infty$. In such limit, the angular uncertainty of the SCS vanishes as compared to the size of the Bloch vector. In that sense, the states are sharply aligned to the coordinate axes of an imagined reference frame. Recently, such a triplet has been introduced as an example of reference frames which separately store non-commuting conserved quantities~\cite{popescu1}. Note that, in general, to construct a QRF for spin one can also use three SCS that are linearly independent but not orthogonal. In this case, the form of the rotation matrix is cumbersome, and so the choice of the orthogonal directions is like a choice of a suitable coordinate system in which the description of physics is simple and convenient.

Let us now return to the situation illustrated in \textit{Figure \ref{fig1}}. According to our construction principle, from the viewpoint of QRF C, the systems are in the state
\begin{equation}
\ket{\Psi}_{A B}^C = \left(\ket{\vec{f}_1}_{A_1}\otimes\ket{\vec{f}_2}_{A_2}\otimes\ket{\vec{f}_3}_{A_3} \otimes \ket{\vec{n}, m}_B\right)^C, \label{state_AB}
\end{equation}
where $\vec{f}_i$, with $i=1,2,3$, name the orientations of the single SCS of A in the QRF~C, and $m$ denotes the spin quantum number associated to the state of B. To this description we could add the three orthonormal vectors $\{\vec{e}_1, \vec{e}_2, \vec{e}_3\}$, which indicate the directions of the three orthogonal axes in the QRF~C. Note however that this is just a notational issue, since these orientations are \emph{fixed} and do not constitute a quantum degree of freedom in the QRF~C. The QRF~C only describes the systems A and B as external systems with dynamical degrees of freedom. Actually, the vectors $\{\vec{e}_1, \vec{e}_2, \vec{e}_3\}$ are going to be always those that indicate the direction of the three orthogonal axes for the reference frame in which we are situated at each moment (which in this case happens to be QRF~C). In particular, they can always be chosen as the ``canonical'' $\vec{e}_1 = (1,0,0), \vec{e}_2 = (0,1,0), \vec{e}_3 = (0,0,1)$.

More generally, the state of a QRF for spin can be given by any quantum superposition or convex mixture of these states with well-defined orientation. Since the extension to mixed states is trivial, we will only consider explicitly pure states. Therefore, in general the systems~A and~B as described in the QRF~C are given by
\begin{equation}
\ket{\Psi}_{A B}^C = \sum_{i, \vec{n}, m} c_{i, \vec{n}, m} \left(\ket{\vec{f}_1^i}_{A_1}\otimes\ket{\vec{f}_2^i}_{A_2}\otimes\ket{\vec{f}_3^i}_{A_3} \ket{\vec{n}, m}_B \right)^C, \label{states_AB}
\end{equation}
where the notation used for the quantum superposition of different SCS of~A and spin states of~B is straightforward. Notice that the reference frame~A and the system~B can be in an entangled state.

Mathematically, the states of A belong to the Hilbert spaces $\mathcal{H}^C_A$ (the superscript denotes the QRF where we are located), where $\mathcal{H}^C_A$ is a tensor products $\mathcal{H}^C_{A_1} \otimes \mathcal{H}^C_{A_2} \otimes \mathcal{H}^C_{A_3}$ of Hilbert spaces of spin-$j$ particles. Each SCS belongs to a corresponding subsystem $\mathcal{H}^C_{A_i}$ of $\mathcal{H}^C_A$, with $i=1,2,3$. The states of system B belong to the Hilbert space $\mathcal{H}^C_B$, which depends on the nature of the spin system B being described. With respect to the QRF~C, the states of the composite system~$A$ and~$B$ belong to the Hilbert space $\mathcal{H}^C_{AB}=\mathcal{H}^C_A\otimes\mathcal{H}^C_B$.

\subsection{Interlude: Classical transformation between reference frames for spin}

Before constructing the transformation between QRF, let us briefly recall the properties of a transformation between two classical reference frames~A and~C, given by the respective sets of orthonormal vectors $\{\vec{f}_1, \vec{f}_2, \vec{f}_3\}$ and $\{\vec{e}_1, \vec{e}_2, \vec{e}_3\}$. The matrix that transforms any vector from its description in the reference frame~C to its description in the reference frame~A is given by
\begin{equation} \label{matrix2}
\mathcal{M}_{C\rightarrow A}=
\left(
\begin{matrix}
\vec{f}_1 \cdot \vec{e}_1 & \vec{f}_1 \cdot \vec{e}_2 & \vec{f}_1 \cdot \vec{e}_3 \\
\vec{f}_2 \cdot \vec{e}_1 & \vec{f}_2 \cdot \vec{e}_2 & \vec{f}_2 \cdot \vec{e}_3 \\
\vec{f}_3 \cdot \vec{e}_1 & \vec{f}_3 \cdot \vec{e}_2 & \vec{f}_3 \cdot \vec{e}_3
\end{matrix}
\right).
\end{equation}

Multiplication by the matrix in~(\ref{matrix2}) transforms any vector~$\vec{n}$, corresponding to the state of system~B, from its description in the classical reference frame~C to its description in the classical reference frame~A. In particular, $\mathcal{M}_{C\rightarrow A} \vec{f}_i = \vec{e}_i$; that is, the vectors $\vec{f}_i$ described in reference frame~C correspond to the three orthogonal axes as described in reference frame~A, as it must be the case. The transformation is therefore consistent with the fact that the vectors $\{\vec{e}_1, \vec{e}_2, \vec{e}_3\}$ are always kept as those providing the three orthogonal axes for the reference frame in which we are situated at each moment. The vectors describing the state of system~B undergo a transformation which gives account of their description in the new reference frame. What remains to be considered is: What is the description now of reference frame~C as described in the reference frame~A? In order to find this, we notice again that the matrix~$\mathcal{M}_{C\rightarrow A}$ (passively) transforms the orthogonal vectors describing reference frame~A to the orthogonal vectors describing reference frame~C. Since in the reference frame~A the orthogonal vectors describing reference frame~A itself are $\{\vec{e}_1, \vec{e}_2, \vec{e}_3\}$, those describing reference frame~C must be $\{\mathcal{M}_{C\rightarrow A} \vec{e}_1, \mathcal{M}_{C\rightarrow A} \vec{e}_2, \mathcal{M}_{C\rightarrow A} \vec{e}_3\}$. In summary, we have that the classical transformation for systems A, B and C is
\begin{equation}
\begin{array}{lrl}
\text{A:} & (\{\vec{f}_1, \vec{f}_2, \vec{f}_3\})^C & \rightarrow (\{\vec{e}_1, \vec{e}_2, \vec{e}_3\})^A, \\
\text{B:} & (\vec{n})^C & \rightarrow (\mathcal{M} \vec{n})^A, \\
\text{C:} & (\{\vec{e}_1, \vec{e}_2, \vec{e}_3\})^C & \rightarrow (\{\mathcal{M} \vec{e}_1, \mathcal{M} \vec{e}_2, \mathcal{M} \vec{e}_3\})^A;
\end{array} \label{transformation}
\end{equation}
where~$\mathcal{M}$ stands for~$\mathcal{M}_{C\rightarrow A}$ in~(\ref{matrix2}), and the superscripts indicate the reference frame in which each system is being described. We notice that the final result of the change of reference frame is simply that all the vectors are rotated with the matrix~$\mathcal{M}_{C\rightarrow A}$, as it should be the case. We have nonetheless carefully discussed the action of the transformation for each system individually since we believe it to be useful in the following.

Since we are considering that the two reference frames share the same chirality, we have that $\det \mathcal{M}_{C\rightarrow A} = 1$, and the transformation can be decomposed in three rotations along different axes, each of the rotations given by an \emph{Euler angle.} In the most common $zxz$-convention for the Euler angles, which we will consider, the first and last rotations are taken along the third axis, while the second one is taken along the second axis. If we call the Euler angles~$\alpha$, $\beta$ and~$\gamma$, these are given in terms of the entries of the matrix~$\mathcal{M}_{C\rightarrow A}$ as
\begin{align}
\alpha & = \sign(\vec{f}_3 \cdot \vec{e}_1)\arccos\left[- \frac{\vec{f}_3 \cdot \vec{e}_2}{\sqrt{1-(\vec{f}_3 \cdot \vec{e}_3)^2}}\right], \nonumber \\
\beta & = \arccos(\vec{f}_3 \cdot \vec{e}_3), \label{classicalbeta} \\
\gamma & = \sign(\vec{f}_1 \cdot \vec{e}_3)\arccos\left[\frac{\vec{f}_2 \cdot \vec{e}_3}{\sqrt{1-(\vec{f}_3 \cdot \vec{e}_3)^2}}\right]; \nonumber
\end{align}
where the ranges considered for the angles are $\alpha \in [-\pi,\pi)$, $\beta \in [0, \pi]$ and $\gamma \in [-\pi,\pi)$, and where one shall consider $\sign(0)=-1$. We give the explicit construction of~$\mathcal{M}_{C\rightarrow A}$ from the three successive rotations in Appendix \ref{euler}. The expressions for~$\alpha$ and~$\gamma$ are not well defined if $\vec{f}_3 \cdot \vec{e}_3 = \pm 1$, what constitutes the so called \emph{gimbal lock,} which is a well-known problem within the transformation between classical reference frames. We also address this problem for QRFs in Appendix \ref{euler}, while from here on we will continue using the expressions above without discussing these not well-defined situations.

\subsection{\label{subsec:level2}Transformation between QRFs for spin}

In quantum mechanics, the operator giving the transformation of a spin state of the system~B, $\ket{\psi}_B$, under a classical reference frame change given by the three Euler angles in~(\ref{classicalbeta}), can be decomposed into three successive rotation operators as $\hat{U}=\rme^{\rmi\gamma\hat{J}^B_{\vec{e}_3}}\cdot \rme^{\rmi\beta\hat{J}^B_{\vec{e}_1}} \cdot \rme^{\rmi\alpha\hat{J}^B_{\vec{e}_3}}$, where $\hat{J}^B_{\vec{n}}$ is the operator corresponding to the component of the angular momentum of the spin system~B along the direction~$\vec{n}$. Notice that the transformation is passive (description from one reference frame to another), hence the change of sign with respect to the active case in the angles.

In the spirit of the construction of QRFs for translational degrees of freedom in~\cite{brukner1}, we will extend this transformation between classical reference frames to QRFs by raising the fixed values for the angles to operators which act on the degrees of freedom of the QRF to which we wish to ``jump''. In order to do so, we will use the expressions in~(\ref{classicalbeta}), just with some minor change, and raise the scalar products~$\vec{f}_i \cdot \vec{e}_j$ that appear in them to operators. For this last task, let us define the following ``cosine operator''~\cite{nienhuis1} acting on the degrees of freedom of one of the orientations of a QRF~K, namely $K_i$:
\begin{equation}
\cos \left( \hat{\theta}^{K_i}_{\vec{l}} \right) := \frac{\hat{J}^{K_i}_{\vec{l}}}{\sqrt{\hat{\vec{J}}^2}} = \frac{\hat{J}^{K_i}_{\vec{l}}}{\hbar\sqrt{j(j+1)}},
\label{angleop}
\end{equation}
where~$\hat{J}^{K_i}_{\vec{l}}$ is the operator corresponding to the component of the angular momentum of the subsystem~$K_i$ along the direction~$\vec{l}$, and where we can write the last equality since we always consider states of total angular momentum~$j$ for the three subsystems of the QRF. If one restricts the cosine function to the interval $\left[0, \pi \right]$ this expression defines the operator~$\hat{\theta}^{K_i}_{\vec{l}}$ uniquely~\cite{nienhuis1}. However, it is the cosine operator itself that we will use. We state the following proposition on this cosine operator acting on SCS states:

\vspace{0.5cm}

\textit{\textbf{Proposition 1:}}
\begin{equation}
\cos \left( \hat{\theta}^{K_i}_{\vec{l}} \right) \ket{\vec{n}}_{K_i} = \vec{n} \cdot \vec{l} \ket{\vec{n}}_{K_i}, \quad \textrm{for}\ j \to \infty. \label{cos}
\end{equation}

\textit{\textbf{Proof:}} We apply the operator~$\hat{J}^{K_i}_{\vec{l}}$ to the SCS in (\ref{spincoherent}). We arbitrarily choose $\vec{l} = (0,0,1)$, so that $\vec{n} \cdot \vec{l} = \cos \theta$. The application of the operator consists in this case just of the multiplication by~$m$ of each term. In the limit $j \rightarrow \infty$, the sum in~$m$ can be approximated by an integral over~$m$, where the binomial distribution is replaced by a Gaussian distribution $\frac{\hbar}{\sigma\sqrt{2\pi}}\rme^{-\frac{1}{2}(\frac{m-\mu}{\sigma})^2}$, with $\sigma=\sqrt{\frac{j}{2}}\sin{\theta}$ and $\mu=j \cos\theta$ (see~\cite{kofler1}). Again, in the limit $j \rightarrow \infty$ this Gaussian tends itself to the Dirac delta $\hbar\ \delta (m - j\cos\theta)$. Therefore, the action of the cosine operator in~(\ref{angleop}) in this limit is given by

\begin{align}
\lim_{j \to \infty} \cos \left( \hat{\theta}^{K_i}_{\vec{l}} \right) \ket{\vec{n}}_{K_i} & = \lim_{j \to \infty} \frac{j \cos\theta}{\sqrt{j(j+1)}} \ket{\vec{n}}_{K_i} \nonumber \\
& = \cos\theta \ket{\vec{n}}_{K_i} = \vec{n} \cdot \vec{l} \ket{\vec{n}}_{K_i} \label{limit}.
\end{align}

\noindent Because of rotational symmetry, this result is valid for any~$\vec{l}$. \hfill $\square$

Having constructed the cosine operator in~(\ref{angleop}), we are in condition to raise the Euler angles in~(\ref{classicalbeta}) to operators, which we do with the following definitions:

\begin{widetext}
\textit{\textbf{Definition 1:} Let $K_1$, $K_2$ and $K_3$ be Hilbert spaces associated to the three directional degrees of freedom of a QRF~K. We introduce the Euler angle operators~$\hat{\alpha}$, $\hat{\beta}$ and~$\hat{\gamma}$ as the operator-valued quantities}
\begin{align}
\hat{\alpha} & := \mathds{1}^{K_1} \otimes \mathds{1}^{K_2} \otimes \frac{2}{\pi} \arctan \left[j \cos \left( \hat{\theta}^{K_3}_{\vec{e_1}} \right) \right] \arccos \left\{- \cos \left( \hat{\theta}^{K_3}_{\vec{e_2}} \right) \left[1-\cos \left( \hat{\theta}^{K_3}_{\vec{e_3}} \right)^2\right]^{-1/2}\right\}, \nonumber\\
\hat{\beta} & :=\mathds{1}^{K_1} \otimes \mathds{1}^{K_2} \otimes \arccos \left[ \cos \left( \hat{\theta}^{K_3}_{\vec{e_3}} \right) \right], \label{beta} \\
\hat{\gamma} & := \frac{2}{\pi} \arctan \left[j \cos \left( \hat{\theta}^{K_1}_{\vec{e_3}} \right) \right] \arccos \left\{\cos \left( \hat{\theta}^{K_2}_{\vec{e_3}} \right) \left[1-\cos \left( \hat{\theta}^{K_3}_{\vec{e_3}} \right)^2\right]^{-1/2}\right\}. \nonumber
\end{align}
\end{widetext}
In these expressions, the analytic functions~$\arctan$ and~$\arccos$ are defined as operator-valued functions in the usual way.

We now prove that the above defined operators~(\ref{beta}) acting on the state of the QRF~A with well-defined orientations given in~(\ref{state_AB}), as described in QRF~C return the classical Euler angles given by~(\ref{classicalbeta}) in the macroscopic limit $j \rightarrow \infty$.

\vspace{0.5cm}

\textit{\textbf{Proposition 2:} The application of the Euler angle operators~(\ref{beta}) on the state $(\ket{\vec{f}_1}_{A_1}\otimes\ket{\vec{f}_2}_{A_2}\otimes\ket{\vec{f}_3}_{A_3})^C$ of the QRF~A as described in the QRF~C with fixed orientations $\{\vec{e}_1, \vec{e}_2, \vec{e}_3\}$ reduces to a multiplication by the classical Euler angles in the limit of large quantum number $j \to \infty$.}

\textit{\textbf{Proof:}} Since all the functions involved in~(\ref{beta}) are analytic for the range of possible values for the operators (except for the mentioned gimbal lock situation, see Appendix~\ref{euler}), it is legitimate to replace the operators in the functions for their eigenvalues in the limit $j \to \infty$ as given in~(\ref{cos}), when they act on SCS. It is therefore quite straightforward to see that the operators in~(\ref{beta}), applied to the corresponding SCS, reduce to the multiplication by the quantities in~(\ref{classicalbeta}) in the limit $j \to \infty$. Only two non-trivial steps require a comment. First, notice that that the function $(2/\pi) \arctan (j x)$ tends to~$\sign x$ in the limit $j \to \infty$ (although is analytic for any finite value of~$j$). Second, notice that because $\cos ( \hat{\theta}^{K_3}_{\vec{l_i}} )$ and $\cos ( \hat{\theta}^{K_3}_{\vec{l_j}} )$ do not commute for $i \neq j$, there is an ambiguity in the definition of $\hat{\alpha}$ regarding the order of the operators. However, since we are working in the limit $j \rightarrow \infty $ when we apply $\hat{\alpha}$ on the SCS, both operators only yield numbers and hence their order becomes irrelevant. \hfill $\square$

\vspace{0.5cm}

It is now rather straightforward to propose the transformation of spin for a change from QRF~C to QRF~A by an unitary transformation~$\hat{\mathcal{S}}_{C \rightarrow A}$. This transformation will consist of three steps. First, we have the transformation of the state of system~B by the following unitary operator:
\begin{equation}
\hat{U}_{C \to A} := \rme^{\rmi\hat{\gamma}\hat{J}^B_{\vec{e}_3}}\cdot \rme^{\rmi\hat{\beta}\hat{J}^B_{\vec{e}_1}} \cdot \rme^{\rmi\hat{\alpha}\hat{J}^B_{\vec{e}_3}},
\label{transf_B}
\end{equation}
where the operators~$\hat{\alpha}$, $\hat{\beta}$ and~$\hat{\gamma}$ are those defined in~(\ref{beta}) and act on the state of~A. Since acting over a state of~A with well-defined orientations (in the limit $j \to \infty$) they consist just of the multiplication by the Euler angle, these operators do not transform such states of~A. Rather, they ``read-out'' the Euler angles so that the operator~$\hat{U}_{C \to A}$ produces the correct rotation on the state of~B, which in general will be different for each term of a quantum superposition (\ref{state_AB}) of states of~A and~B.

Second, we need to transform the states of the QRF. Notice that, since we limit the possible states of the QRF to SCS in the limit of infinitely large spin, we do not need to explicitly construct an operator that acts on any possible quantum state. Rather, we only need to operationally define how this operator acts on such family of states. Also notice that we need to change the state of the QRF from describing the state of~A in the QRF~C to describing the state of~C in the QRF~A. Recalling the classical transformation in~(\ref{transformation}), we see that the action of the operator on the state describing the QRF must be
\begin{multline}
\hat{V}_{C \to A} \left(\ket{\vec{f}_1}_{A_1}\otimes\ket{\vec{f}_2}_{A_2}\otimes\ket{\vec{f}_3}_{A_3} \otimes \ket{\psi}_B\right)^C = \\
\left(\ket{\mathcal{M} \vec{e}_1}_{A_1}\otimes\ket{\mathcal{M} \vec{e}_2}_{A_2}\otimes\ket{\mathcal{M} \vec{e}_3}_{A_3} \otimes \ket{\psi}_B \right)^C.
\label{transf_C}
\end{multline}
In this expression, the operator needs to ``read out'' the state of~A in the QRF~C in order to implement the correct rotation~$\mathcal{M}_{C\rightarrow A}$, which depends on the vectors~$\vec{f}_i$, to the vectors~$\vec{e}_i$. Of course, the way this reading of the entries of~$\mathcal{M}_{C\rightarrow A}$ can be accomplished is, as for the entries appearing in the Euler angles, through the use of the operator~(\ref{angleop}) and its property~(\ref{cos}), although for the lack of simplicity we have not written down this step explicitly.

The third step of the transformation is trivial and consists only of an ``operator'' $\hat{\mathcal{P}}_{A \leftrightarrow C}$ that changes the labels $A \leftrightarrow C$ of the two QRFs. Collecting all three steps, the operator implementing the transformation from QRF~C to QRF~A is
\begin{equation}
\hat{\mathcal{S}}_{C \to A} := \hat{\mathcal{P}}_{A \leftrightarrow C} \hat{V}_{C \to A} \hat{U}_{C \to A}.
\label{trafo}
\end{equation}

Let us check that this operator acts as expected on a state with well defined orientations such as that in~(\ref{state_AB}):
\begin{widetext}
\begin{align}
\hat{\mathcal{S}}_{C \to A} \ket{\Psi}_{A B}^C & = \hat{\mathcal{S}}_{C \to A} \left(\ket{\vec{f}_1}_{A_1}\otimes\ket{\vec{f}_2}_{A_2}\otimes\ket{\vec{f}_3}_{A_3} \otimes \ket{\vec{n}, m}_B\right)^C \nonumber \\
& = \hat{\mathcal{P}}_{A \leftrightarrow C} \hat{V}_{C \to A} \hat{U}_{C \to A} \left(\ket{\vec{f}_1}_{A_1}\otimes\ket{\vec{f}_2}_{A_2}\otimes\ket{\vec{f}_3}_{A_3} \otimes \ket{\vec{n}, m}_B\right)^C \nonumber \\
& = \hat{\mathcal{P}}_{A \leftrightarrow C} \hat{V}_{C \to A} \rme^{\rmi\hat{\gamma}\hat{J}^B_{\vec{e}_3}}\cdot \rme^{\rmi\hat{\beta}\hat{J}^B_{\vec{e}_1}} \cdot \rme^{\rmi\hat{\alpha}\hat{J}^B_{\vec{e}_3}} \left(\ket{\vec{f}_1}_{A_1}\otimes\ket{\vec{f}_2}_{A_2}\otimes\ket{\vec{f}_3}_{A_3} \otimes \ket{\vec{n}, m}_B\right)^C \nonumber \\
& = \hat{\mathcal{P}}_{A \leftrightarrow C} \hat{V}_{C \to A} \rme^{\rmi \gamma \hat{J}^B_{\vec{e}_3}}\cdot \rme^{\rmi \beta \hat{J}^B_{\vec{e}_1}} \cdot \rme^{\rmi \alpha \hat{J}^B_{\vec{e}_3}} \left(\ket{\vec{f}_1}_{A_1}\otimes\ket{\vec{f}_2}_{A_2}\otimes\ket{\vec{f}_3}_{A_3} \otimes \ket{\vec{n}, m}_B\right)^C \nonumber \\
& = \hat{\mathcal{P}}_{A \leftrightarrow C} \hat{V}_{C \to A} \left(\ket{\vec{f}_1}_{A_1}\otimes\ket{\vec{f}_2}_{A_2}\otimes\ket{\vec{f}_3}_{A_3} \otimes \ket{\mathcal{M}_{C\rightarrow A} \vec{n}, m}_B\right)^C \nonumber \\
& = \hat{\mathcal{P}}_{A \leftrightarrow C} \left(\ket{\mathcal{M}_{C\rightarrow A} \vec{e}_1}_{A_1}\otimes\ket{\mathcal{M}_{C\rightarrow A} \vec{e}_2}_{A_2}\otimes\ket{\mathcal{M}_{C\rightarrow A} \vec{e}_3}_{A_3} \otimes \ket{\mathcal{M}_{C\rightarrow A} \vec{n}, m}_B\right)^C \nonumber \\
& = \left(\ket{\mathcal{M}_{C\rightarrow A} \vec{e}_1}_{C_1}\otimes\ket{\mathcal{M}_{C\rightarrow A} \vec{e}_2}_{C_2}\otimes\ket{\mathcal{M}_{C\rightarrow A} \vec{e}_3}_{C_3} \otimes \ket{\mathcal{M}_{C\rightarrow A} \vec{n}, m}_B\right)^A.
\label{transf_check}
\end{align}
\end{widetext}
We notice that, in the fourth step of the derivation, the Euler angle operators simply got ``replaced'' by their corresponding values depending on the state of the QRF~A, since it is by construction an eigenstate of all three operators (in the limit $j \to \infty$).

By linearity of the operator, we can apply it also to any general state of a QRF~A and a system~B, such as the quantum superpositions in~(\ref{states_AB}). After the transformation, the three vectors $\{\vec{e}_1, \vec{e}_2, \vec{e}_3\}$ become the directions in space for the QRF~A, and the state of~A is not considered as a degree of freedom any more~\footnote{We notice again that each of the three rotations with the respective Euler angles are \emph{passive} transformation. This means that every time we perform one of the rotations, we also change the reference frame. In our convention, this reference frame is always by definition called $\{\vec{e}_1, \vec{e}_2, \vec{e}_3\}$, and thus when the process finishes these three vectors have changed from denoting the three directions in space of QRF~C to denoting the three directions in space of QRF~A. But conceptually it is important to notice that, for example, the second rotation along the $\vec{e}_1$-axis is not taken along such original axis of QRF-C, but along the new axis after the first rotation has already taken place. This is of course correctly implemented by the expressions given in~(\ref{transf_check}), since they correspond to passive transformations.}.

\subsection{\label{sec:invar} Invariance of Hamiltonians under quantum reference frame transformations}

In physics, a transformation that leaves the Hamiltonian invariant is called a symmetry. When we describe dynamics from QRFs, the Hamiltonian as seen from one QRF includes not only the system B but also the other QRF A. In Ref~\cite{brukner1}, a symmetry was defined as a mapping that leaves the functional form of the Hamiltonian invariant, i.e. the Hamiltonian of A and B is the same function of operators as the Hamiltonian of C and B. Consider now the set of Hamiltonians for spin quantum systems which are invariant under rotations of (classical) reference frames. We will show that they are also invariant under a larger group of the QRF transformations~(\ref{trafo}).

We consider a Hamiltonian between the system~B and the QRF~A, as described in the QRF~C, which is invariant under common identical rotations on QRF~A and system~B. This implies that:
\begin{equation}
\hat{R} \hat{H}^{C}_{AB} \hat{R}^\dagger = \hat{H}^{C}_{AB}, \quad \hat{R} := \hat{M}_{A_1} \otimes \hat{M}_{A_2} \otimes \hat{M}_{A_3} \otimes \hat{M}_B
\label{HamiltonianAB}
\end{equation}
where $\hat{M}$ is a general representation of \emph{the same} rotation on each subsystem. The Hamiltonian can contain the Heisenberg interaction (for example if we were to introduce two spin systems B$_1$ and B$_2$) or an interaction terms between the QRF~A and the system B. The latter case, due to the semi-classical nature of the states for the QRF, would correspond to an interaction between spin system B and an external field, which could be in a quantum superposition of different orientations.

In Appendix~\ref{Heisenberg} we prove the invariance of~(\ref{HamiltonianAB}) under a change of QRF. That is, when we change to the description in QRF~A, the Hamiltonian preserves its form, only changing the system labels:
\begin{align}
\hat{H}^{A}_{B C} = \hat{\mathcal{P}}_{A \leftrightarrow C} \hat{H}^{C}_{A B} \hat{\mathcal{P}}_{A \leftrightarrow C}.
\label{HamiltonianBC}
\end{align}
This shows that the Hamiltonian has a larger symmetry than rotational invariance. Moreover, the Schr\"odinger equation with the Hamiltonian exhibits generalized covariance of dynamical laws~\cite{brukner1}, since the laws preserve their form not only under rotations but also under ``superposition of rotations''. We shall remark however that, in proving the invariance, we do not consider the action of the Hamiltonian on the entire Hilbert space ${\cal H}^C_{AB}$ but only onto the semi-classical states of the form~(\ref{states_AB}). Finally, let us notice that generalizing the result for rotationally invariant Hamiltonians for more spin quantum systems is clearly straightforward.

\section{\label{subsec:level3}Examples}

Let us now continue with the application of transformation (\ref{trafo}) to some physically relevant states $\ket{\psi}_{AB}^C$. We choose B to be a spin-1/2-particle and give three examples of representative initial states and their transformed counterparts.

\paragraph{\label{ex1}Rotated QRFs}

\begin{figure}[t]
\centering
\includegraphics[width=0.6\linewidth]{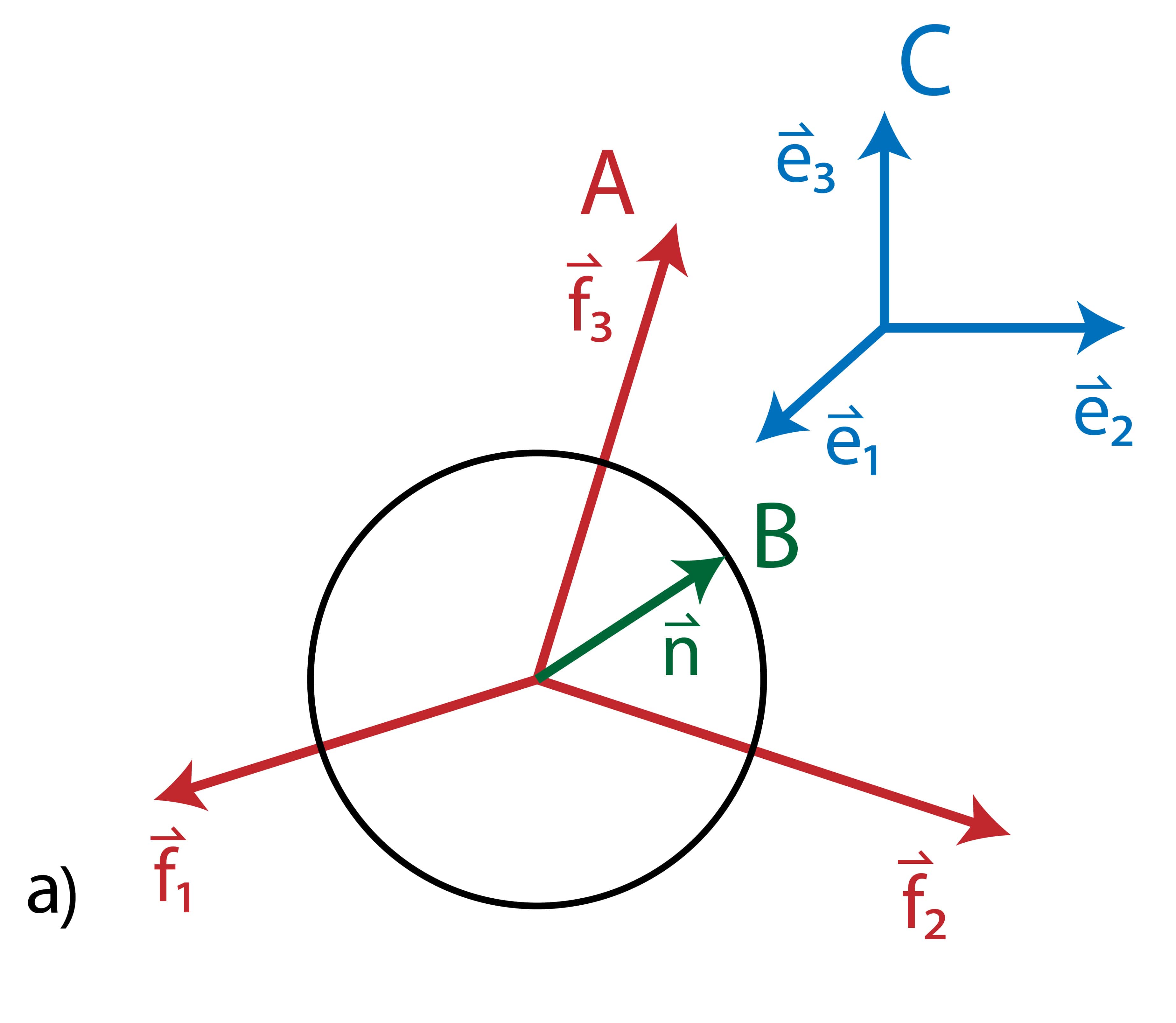}
\includegraphics[width=0.6\linewidth]{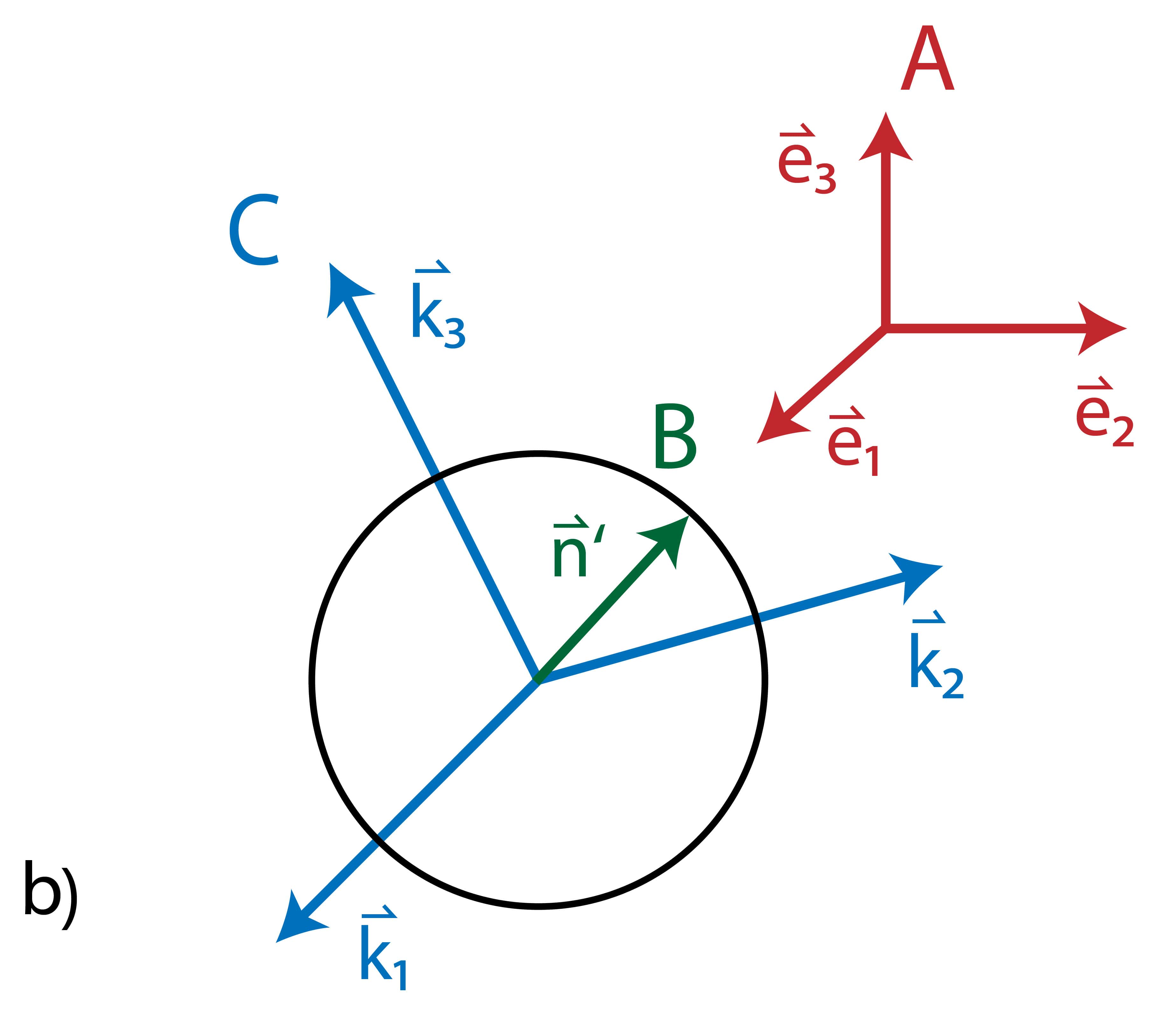}
\caption{Illustration of Example a: Rotated QRFs \\
a) The SCS of QRF A form a right-handed system whose axes are oriented along arbitrary directions $\vec{f}_1, \vec{f}_2$ and $\vec{f}_3$ with respect to QRF C in the limit $j\rightarrow \infty$. The spin-$1/2$-particle B is depicted by a general Bloch state $\vec{n}$.
b) After the transformation to QRF A, both C's SCS and B's spin-$1/2$-state possess new orientations $\vec{k}_1, \vec{k}_2, \vec{k}_3$ and $\vec{n}'$ in A.
\label{fig3}}
\end{figure}

Let us first look at the case in which A's and C's SCS are just rotated with respect to each other, analogously to two classical reference frames with right-handed bases $\{\vec{e}_1,\vec{e}_2,\vec{e}_3\}$ and $\{\vec{f}_1,\vec{f}_2,\vec{f}_3\}$, respectively (see \textit{Figure~\ref{fig3}}) . In QRF C, A and B are given by the state
\begin{align}
\ket{\psi}_{AB}^C&=
\left( \ket{\vec{f}_1}_{A_1}
\otimes \ket{\vec{f}_2}_{A_2}
\otimes \ket{\vec{f}_3}_{A_3}
\otimes \ket{\vec{n},1/2}_B
\right)^C,
\end{align}
where $\vec{n}=
\left( \begin{matrix}
\sin\theta\cos\phi\\
\sin\theta\sin\phi\\
\cos\theta
\end{matrix}\right)$
describes an arbitrary spin-1/2-state parameterized by two angles $\theta$ and $\phi$. After the transformation to QRF A, the new state of B and C seen in A is given by
\begin{align}
\ket{\psi}_{CB}^A&=
\left( \ket{\vec{k}_1}_{C_1}
\otimes \ket{\vec{k}_2}_{C_2}
\otimes \ket{\vec{k}_3}_{C_3}
\otimes \ket{\vec{n}',1/2}_B
\right)^A
\end{align}
with C's new SCS orientations
\begin{align}
\vec{k}_1&=
\mathcal{M}_{C \rightarrow A}\vec{e}_1=
\left(
\begin{matrix}
\cos\alpha\cos\gamma-\sin\alpha\cos\beta\sin\gamma\\
-\cos\alpha\sin\gamma-\sin\alpha\cos\beta\cos\gamma \\ \sin\alpha\sin\beta
\end{matrix}
\right), \\
~~~~~\vec{k}_2&=
\mathcal{M}_{C \rightarrow A}\vec{e}_2=
\left(
\begin{matrix}
\sin\alpha\cos\gamma+\cos\alpha\cos\beta\sin\gamma\\
-\sin\alpha\sin\gamma+\cos\alpha\cos\beta\cos\gamma \\ -\cos\alpha\sin\beta
\end{matrix}
\right),\\
\vec{k}_3&=
\mathcal{M}_{C \rightarrow A}\vec{e}_3=
\left(
\begin{matrix}
\sin\beta\sin\gamma\\
\sin\beta\cos\gamma \\
\cos\beta
\end{matrix}
\right),
\end{align}
and the new spin-1/2-orientation
\begin{widetext}
\begin{align}
&\vec{n}'=
\left(
\begin{matrix}
\sin \beta \sin \gamma \cos \theta+\sin \theta[\cos (\alpha+\gamma-\phi)\cos^2(\frac{\beta}{2})+\cos(\alpha-\gamma-\phi) \sin^2(\frac{\beta}{2})]\\
\sin \beta \cos\gamma \cos \theta - \sin \theta[ \sin (\alpha+\gamma-\phi) \cos^2(\frac{\beta}{2})- \sin (\alpha-\gamma-\phi) \sin^2(\frac{\beta}{2})]\\
\cos\beta \cos\theta + \sin\theta \sin(\alpha-\phi) \sin\beta
\end{matrix}
\right)
\end{align}
\end{widetext}
in QRF A. Here, $\alpha$, $\beta$ and $\gamma$ name the three Euler angles ~(\ref{classicalbeta}) occurring in the QRF transformation.

\begin{figure}
\includegraphics[width=0.8\linewidth]{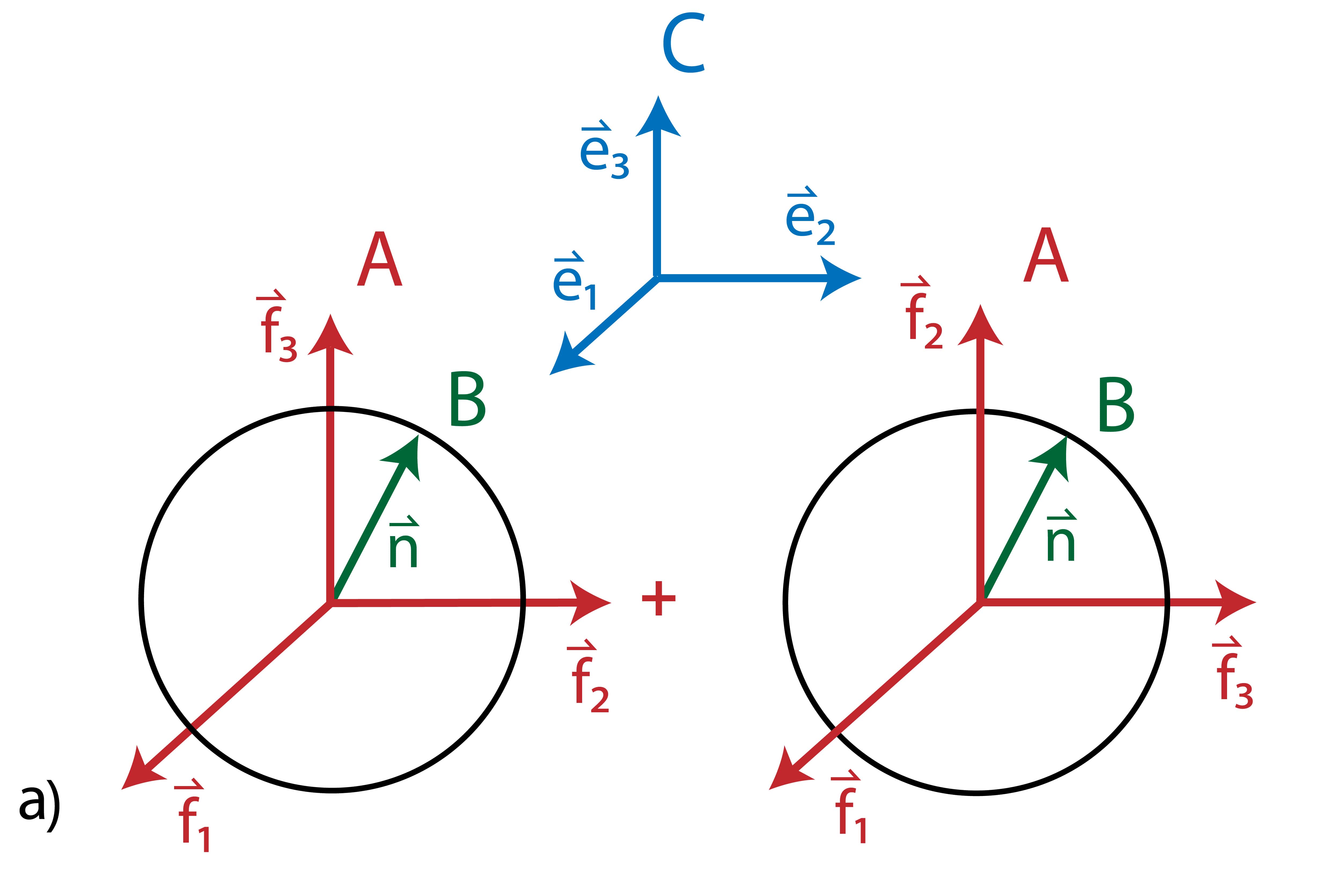}
\includegraphics[width=0.8\linewidth]{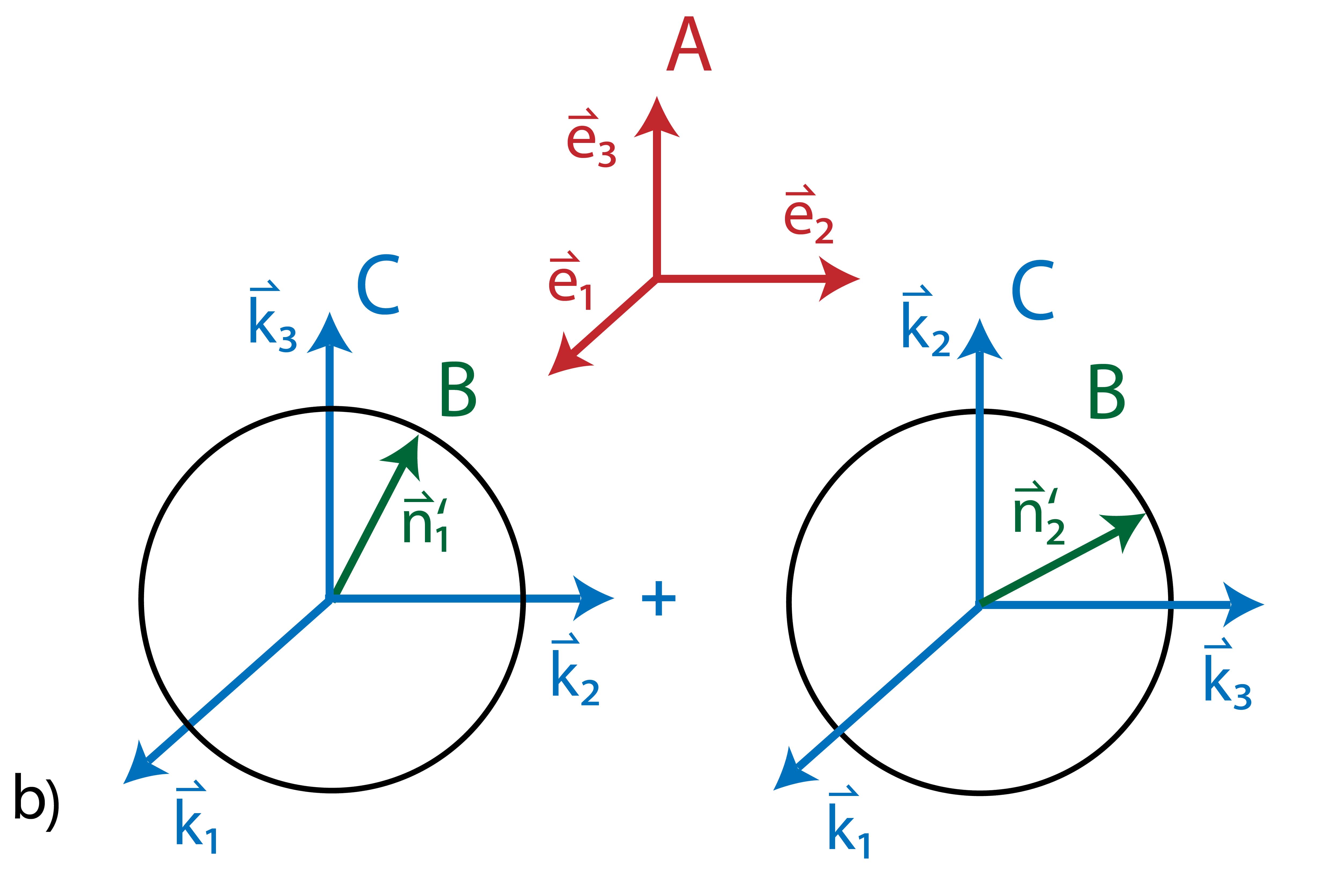}%
\caption{Illustration of Example b: Superposed QRF \\
a) QRF A is in a superposition of two amplitudes corresponding to two oppositely handed frames with respect to QRF C. The spin-$1/2$-particle B is described by an arbitrarily oriented Bloch vector $\vec{n}$.
b) Switching perspective to QRF A, the state of system $C\otimes B$ becomes an entangled state.\label{fig4}}
\end{figure}

\paragraph{\label{ex2}Superposed QRF}

We continue with the case where QRF A is a superposition with respect to QRF C, such that the right-handed amplitude of A and frame C are equally oriented, while for the left-handed amplitude of A the $x$ axis coincides with the one of C, but their $y$ and $z$ axes are swapped (see \textit{Figure~\ref{fig4}}). Such a state is given by
\begin{align}
\ket{\psi}_{AB}^C =&
\frac{1}{\sqrt{2}}
\left[\left(
\ket{\vec{e}_1}_{A_1}
\otimes\ket{\vec{e}_2}_{A_2}
\otimes\ket{\vec{e}_3}_{A_3}
\right.\right.\nonumber \\
&+ \left.\left.\rme^{\rmi\Phi}
\ket{\vec{e}_1}_{A_1}
\otimes\ket{\vec{e}_3}_{A_2}
\otimes\ket{\vec{e}_2}_{A_3}\right)
\otimes\ket{\vec{n},1/2}_B
\right]^C,
\end{align}
with A's axes $\{\vec{f}_1,\vec{f}_2,\vec{f}_3\}$ oriented along $\{\vec{e}_1,\vec{e}_2,\vec{e}_3\}$ and $\{\vec{e}_1,\vec{e}_3,\vec{e}_2\}$ for the two superposed amplitudes, respectively, and $\rme^{\rmi\Phi}$ is an arbitrary complex phase. In this case, the Euler angles $(\alpha,\beta,\gamma)$ are $(0,0,0)$ and $(-\pi,\frac{\pi}{2},0)$ for the two superposed orientations of A with respect to C. However, the second amplitude of QRF A is oppositely handed to QRF C and therefore experiences also a reflection during the transformation (see Appendix~\ref{euler}). The corresponding matrices $\mathcal{M}_{C\rightarrow A}$ and $\mathcal{M}^{\rm{impr}}_{C\rightarrow A}$ are given by
\begin{align}
\mathcal{M}_{C\rightarrow A} (0,0,0)&=\mathds{1}, \\
\mathcal{M}^{\rm{impr}}_{C\rightarrow A}(-\pi,\frac{\pi}{2},0)&=
\left(
\begin{matrix}
1 & 0 & 0 \\
0 & 0 & 1 \\
0 & 1 & 0
\end{matrix}
\right).
\end{align}
After the change of perspective, in QRF A system B appears in an entangled state with QRF C:
\begin{align}
\ket{\psi}_{CB}^A =&
\frac{1}{\sqrt{2}}
\left[\left(
\ket{\vec{e}_1}_{C_1}
\otimes\ket{\vec{e}_2}_{C_2}
\otimes\ket{\vec{e}_3}_{C_3}
\otimes\ket{\vec{n}_1',1/2}_B
\right.\right.\nonumber \\
&+ \left.\left.\rme^{\rmi\Phi}
\ket{\vec{e}_1}_{C_1}
\otimes\ket{\vec{e}_3}_{C_2}
\otimes\ket{\vec{e}_2}_{C_3}\right)
\otimes\ket{\vec{n}_2',1/2}_B
\right]^A,
\end{align}
Here,
$\vec{n}_1'=\left(
\begin{matrix}
\sin\theta\cos\phi\\
\sin\theta\sin\phi\\
\cos\theta
\end{matrix}
\right)$
and
$\vec{n}_2'=
\left(
\begin{matrix}\sin\theta\cos\phi\\
\cos\theta\
\\\sin\theta\sin\phi
\end{matrix}
\right)$
are the new spin-1/2-orientations in A and C's axes $\{\vec{k}_1,\vec{k}_2,\vec{k}_3\}$ are oriented along $\{\vec{e}_1,\vec{e}_2,\vec{e}_3\}$ and $\{\vec{e}_1,\vec{e}_3,\vec{e}_2\}$, respectively.

This example illustrates the fact that superposition and entanglement are QRF-dependent notions. In particular, in this simple example we can see that, when moving to a QRF C in a coherent superposition of different orientations, but not entangled with the system B being described, one finds that, from the perspective of QRF C, the original QRF A is entangled with system B. The degree of entanglement depends however on the concrete coherent superposition of QRF C, and there situations for which no entanglement is created, for example if QRF C is in a superposition of orientations which are all related by rotations along the axis of $\vec{n}$. Quantitatively, the entanglement also depends of course on the nature and state of system B. As we mentioned, this ``conversion relation'' between coherence and entanglement under QRF transformations is a feature appearing also for translational degrees of freedom~\cite{brukner1}. Studying in more depth such relation between these two quantities, in light of works which relate them such as \cite{PhysRevLett.115.020403}, will be subject of future research by the authors.

\begin{figure}
\includegraphics[width=0.8\linewidth]{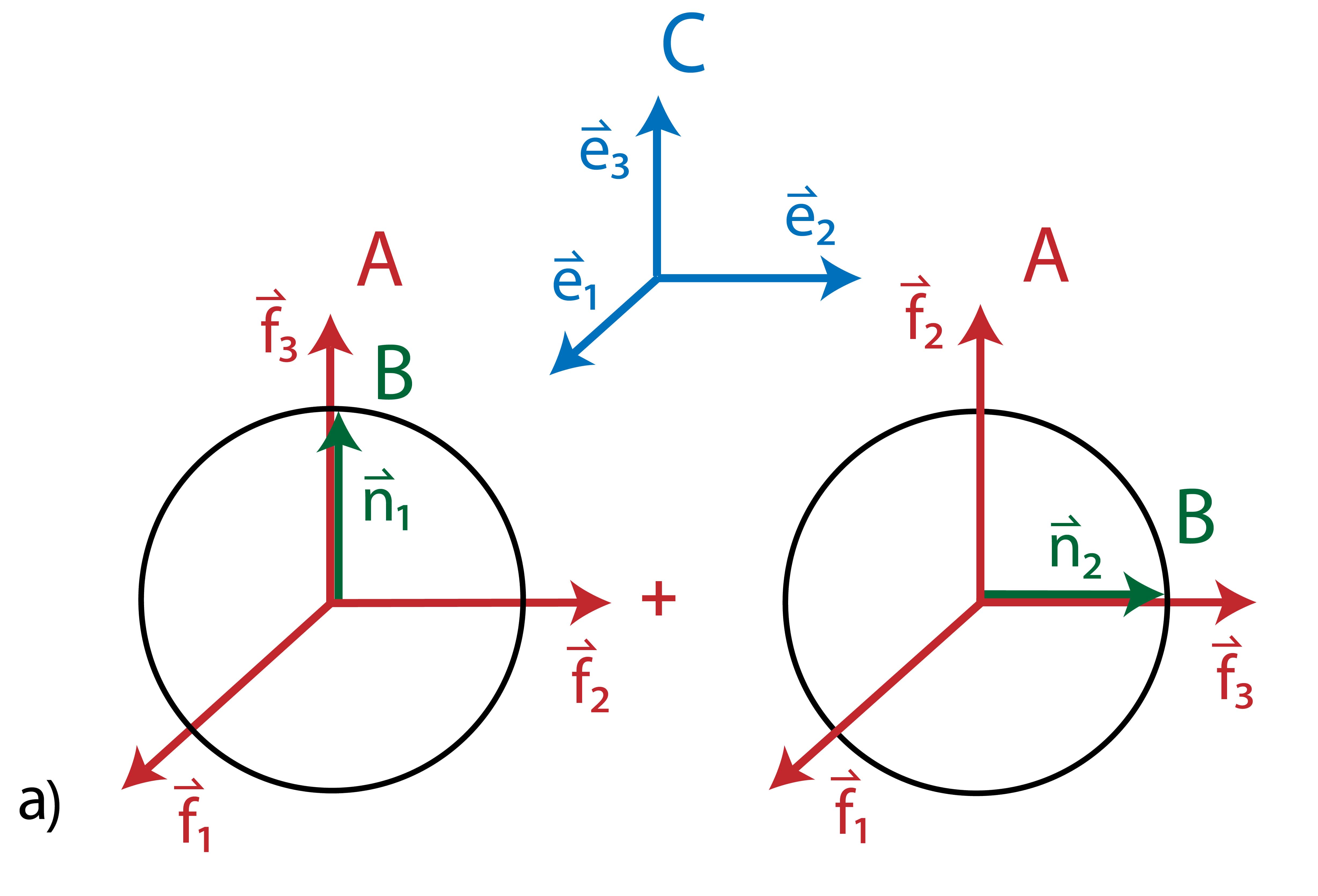}
\includegraphics[width=0.8\linewidth]{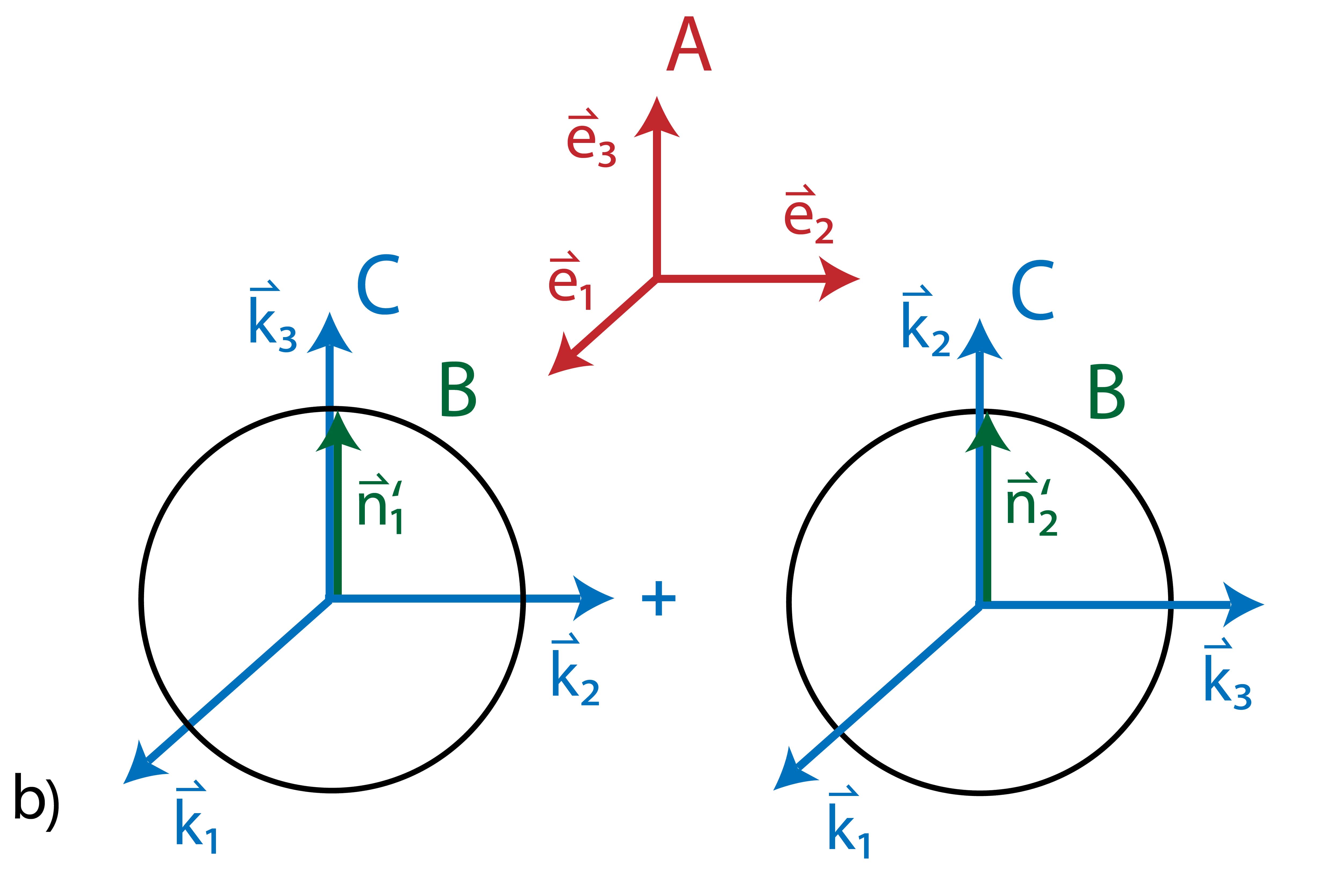}%
\caption{Illustration of Example c: Entangled QRF\\
a) QRF A and spin-1/2-particle B are in an entangled state in which the two different orientations of B are aligned to the $z$ axes of two oppositely handed frames A in QRF C.
b) As expected, after a jump to QRF A, B reduces to a pure state with orientation along A's $z$ axis, i.e. $\vec{n}'_1=\vec{n}'_2=\vec{e}_3$, leaving only C in a superposition state. Here, it becomes apparent, that superposition and entanglement are frame-dependent concepts. \label{fig5}}
\end{figure}

\paragraph{\label{ex3}Entangled QRF}

Let us finally consider the opposite case in which A and B form an entangled state in QRF C. We choose two amplitudes of A in the entangled state to correspond to two oppositely handed frames. These states are correlated to spin states of B, which are aligned along the
$z$ axes of the two frames of A (see \textit{Figure~\ref{fig5}}):
\begin{align}
\ket{\psi}_{AB}^C =&
\frac{1}{\sqrt{2}}\left[
\ket{\vec{e}_1}_{A_1}\otimes
\ket{\vec{e}_2}_{A_2}\otimes
\ket{\vec{e}_3}_{A_3}\otimes
\ket{\vec{e}_3,1/2}_B)
\right.\nonumber \\
&+\left.\rme^{\rmi\Phi}
\ket{\vec{e}_1}_{A_1}\otimes
\ket{\vec{e}_3}_{A_2}\otimes
\ket{\vec{e}_2}_{A_3}\otimes
\ket{\vec{e}_2,1/2}_B\right] ^C.
\end{align}
After the transformation to frame QRF A's perspective, the spin-1/2-state factorizes and only C appears to be in a superposition state
\begin{align}
\ket{\psi}_{CB}^A =&
\frac{1}{\sqrt{2}}\left[
\left(\ket{\vec{e}_1}_{C_1}\otimes
\ket{\vec{e}_2}_{C_2}\otimes
\ket{\vec{e}_3}_{C_3} \right.\right.\nonumber \\
&+\left.\left.\rme^{\rmi\Phi}
\ket{\vec{e}_1}_{C_1}\otimes
\ket{\vec{e}_3}_{C_2}\otimes
\ket{\vec{e}_2}_{C_3}\right)
\otimes\ket{\vec{e}_3,1/2}_B\right]^A.
\end{align}

\section{\label{sec:level4}Discussion}

The formalism of QRFs has been implemented within the framework of quantum mechanics and features an approach that is categorically relational by construction, i.e. all physical information is encoded in relative variables.

In this work we extend the formalism of QRFs~\cite{brukner1} to spin, and establish a method to promote certain spin quantum systems to reference frames for spin. We propose the construction of such ``quantized'' spin frames by compositions of three spin coherent states. While limited to systems composed of three orthogonal SCS with infinitely large spin, the resulting quantized description enables the consideration of QRFs whose spin degrees of freedom are in a superposition or an entangled states.

We find unitary transformations that map quantum descriptions from different QRFs. To this end, we follow an operational approach uplifting the Euler angles that occur in a classical spin transformation mathematically to the rank of operators. This procedure preserves the quantum nature of the frames and adds a direct physical meaning to the switch of perspective between them.

We consider dynamics of spin systems from different QRFs and find the Hamiltonian under common rotations of the classical reference frame to be invariant under the change of QRFs. This extends the group of symmetry of these Hamiltonian beyond the rotational invariance. The invariance might be of practical importance when one wants to determine how spin evolves in an external magnetic field that is in a non-classical state. The form of the Hamiltonian is preserved when one ``jumps'' to the QRF of the magnetic field.

Finally, we apply our formalism to three physically interesting cases, including superposition states as well as entangled ones; and consequently find that, analogously to position and momentum variables~\cite{brukner1}, spin superposition and entanglement are observer-dependent notions.

We comment on possible extensions of our work. First, the techniques introduced here together with the formalism provided in Refs.~\cite{brukner1,brukner2} could be used to describe measurement procedures including Stern-Gerlach apparatuses. Instead of being considered to have a fixed orientation with respect to the laboratory reference frame, the apparatus can be prepared in a superposition therefrom.

Second, our methods works in the semi-classical limit of large angular momentum in which the Euler angle operators have continuous spectra~\cite{nienhuis1}. It would be interesting to go beyond infinite spins and develop a formalism for QRFs with finite dimensions in which the mathematical structure of the underlying Hilbert space is fundamentally different. In this case, the quantized transformation might not be unitary~\cite{hamette1}. This would have far-reaching consequences for the description of physical realities, since the choice of a reference frame would accordingly impact the evolution of the involved systems itself.

\vspace{0.5cm}

\begin{acknowledgments}

The authors want to thank Esteban Castro, Thomas D. Galley, Flaminia Giacomini and Anne-Catherine de la Hamette; and an anonymous referee for his/her constructive criticism and nice suggestions, which greatly helped to improve the article. L.C.B.\ acknowledges the support from the research platform TURIS, from the \"{O}AW through the project ``Quantum Reference Frames for Quantum Fields'' (ref.~IF\textunderscore 2019\textunderscore 59\textunderscore QRFQF) and from the European Commission via Testing the Large-Scale Limit of Quantum Mechanics (TEQ) (No. 766900) project. The authors were also supported by the Austrian-Serbian bilateral scientific cooperation no. 451-03-02141/2017-09/02, and by the Austrian Science Fund (FWF) through the SFB project BeyondC (sub-project~F7103) and a grant from the Foundational Questions Institute (FQXi) Fund. This publication was made possible through the support of the ID61466 grant from the John Templeton Foundation, as part of the The Quantum Information Structure of Spacetime (QISS) Project (qiss.fr). The opinions expressed in this publication are those of the author(s)and do not necessarily reflect the views of the John Templeton Foundation.

\end{acknowledgments}

\appendix

\begin{widetext}

\section{Rotation matrix and special cases}\label{euler}

For transformations between reference frames with the same chirality, the construction of the matrix~$\mathcal{M}_{C\rightarrow A}$ reads

\begin{equation}
\mathcal{M}_{C\rightarrow A} =
\left( \begin{array}{ccc}
	\cos \gamma & \sin \gamma & \\
	-\sin \gamma & \cos \gamma & \\
	 & & 1
\end{array}
\right)
\left( \begin{array}{ccc}
	1 & & \\
	& \cos \beta & \sin \beta \\
	& -\sin \beta & \cos \beta
\end{array}
\right)
\left( \begin{array}{ccc}
	\cos \alpha & \sin \alpha & \\
	-\sin \alpha & \cos \alpha & \\
	 & & 1
\end{array}
\right).
\label{matrix_rotations}
\end{equation}

\end{widetext}

Let us consider the case of reference frames with opposite chirality, which clearly requires an \emph{improper} orthogonal transformation. The dependences of the Euler angles with respect to the matrix entries, given in~(\ref{classicalbeta}), are obtained from the components of the axes~$\vec{f}_2$ and~$\vec{f}_3$ of reference frame~A in the old reference frame~C (see~\cite{wiki:001, goldstein:mechanics}); \emph{except} for the sign of~$\gamma$, which is determined by the third component of~$\vec{f}_1$. Clearly, if we introduce a reflection along the ${\vec{e}_1}$-direction between the second and the third rotation, this will produce a reference frame with opposite chirality, while the dependence of the Euler angles with respect to the matrix entries will not be affected (since such reflection does not modify~$\vec{f}_2$ or~$\vec{f}_3$) \emph{except} for a change of the sign of~$\gamma$. But, since the rotation with the angle~$\gamma$ is taken \emph{after} the reflection, this fact also flips the sense in which rotations with positive and negative angles must be considered, and compensates for the sign flip. Therefore, if we use the same definitions for the Euler angles in~(\ref{classicalbeta}) also for the case of opposite chirality, the transformation matrix would read

\begin{widetext}

\begin{equation}
\mathcal{M}^{\rm{impr}}_{C\rightarrow A} =
\left( \begin{array}{ccc}
	\cos \gamma & \sin \gamma & \\
	-\sin \gamma & \cos \gamma & \\
	 & & 1
\end{array}
\right)
\left( \begin{array}{ccc}
	-1 & & \\
	& \cos \beta & \sin \beta \\
	& -\sin \beta & \cos \beta
\end{array}
\right)
\left( \begin{array}{ccc}
	\cos \alpha & \sin \alpha & \\
	-\sin \alpha & \cos \alpha & \\
	 & & 1
\end{array}
\right),
\label{matrix_impr}
\end{equation}

\end{widetext}

\noindent where the reflection along the ${\vec{e}_1}$-direction is introduced with the ``-1'' element of the second matrix. Accordingly, the most general quantum transformation is given by
\begin{equation}
\hat{U}_{C \to A} := \rme^{\rmi\hat{\gamma}\hat{J}^B_{\vec{e}_3}} \cdot \hat{R}_{\vec{e}_1}^{(1-|\hat{\mathcal{M}}|)/2} \cdot \rme^{\rmi\hat{\beta}\hat{J}^B_{\vec{e}_1}} \cdot \rme^{\rmi\hat{\alpha}\hat{J}^B_{\vec{e}_3}};
\label{transf_B_general}
\end{equation}
where~$|\hat{\mathcal{M}}|$ is an operator reading out the determinant of the matrix~$\mathcal{M}_{C\rightarrow A}$ as defined in~(\ref{matrix2}) for each quantum state of the reference frame~A, and~$ \hat{R}_{\vec{e}_1}$ is the reflection along the $\vec{e}_1$-axis.

As for the problem with the gimbal lock, we shall give explicit alternative expressions for the Euler angles when $\vec{f}_3 \cdot \vec{e}_3 = \pm 1$. One possibility is
\begin{align}
\alpha & = 0, \nonumber \\
\beta & = (1 - \vec{f}_3 \cdot \vec{e}_3) \pi/2, \label{gimbal} \\
\gamma & = \sign[(\vec{f}_3 \cdot \vec{e}_3) (\vec{f}_1 \cdot \vec{e}_2)]\arccos[ (\vec{f}_3 \cdot \vec{e}_3) (\vec{f}_2 \cdot \vec{e}_2)].
\end{align}
These expressions can be used in the general definitions of the rotation operations for both chiralities.

\section{Proof of the invariance of the Hamiltonian under QRF transformations}\label{Heisenberg}

We consider the transformed Hamiltonian $\hat{H}^A_{BC} = \hat{\mathcal{S}} \hat{H}^C_{AB} \hat{\mathcal{S}}^\dagger$ and let it act on SCS of QRF~C and spin states of B:

\begin{align}
\hat{H}^A_{BC} & |\vec{g}_1\rangle_{C_1} \otimes |\vec{g}_2\rangle_{C_2} \otimes |\vec{g}_3\rangle_{C_3} \otimes \ket{\vec{m}}_B \nonumber \\
& = \hat{\mathcal{S}} \hat{H}^C_{AB} \hat{\mathcal{S}}^\dagger |\vec{g}_1\rangle_{C_1} \otimes |\vec{g}_2\rangle_{C_2} \otimes |\vec{g}_3\rangle_{C_3} \otimes \ket{\vec{m}}_B \nonumber \\
& = \hat{\mathcal{S}} \hat{H}^C_{AB} \hat{U}_{C \to A}^\dagger \hat{V}_{C \to A}^\dagger \hat{\mathcal{P}}_{A \leftrightarrow C} |\vec{g}_1, \vec{g}_2, \vec{g}_3, \vec{m} \rangle_{BC} \nonumber \\
& = \hat{\mathcal{S}} \hat{H}^C_{AB} \hat{U}_{C \to A}^\dagger \hat{V}_{C \to A}^\dagger |\vec{g}_1, \vec{g}_2, \vec{g}_3, \vec{m} \rangle_{AB} \nonumber \\
& = \hat{\mathcal{S}} \hat{H}^C_{AB} |{\cal M}^{-1}\vec{g}_1,{\cal M}^{-1}\vec{g}_2,{\cal M}^{-1}\vec{g}_3,{\cal M}^{-1}\vec{m} \rangle_{AB} \nonumber \\
& = \hat{\mathcal{S}} \hat{H}^C_{AB} \hat{R}^\dagger |\vec{g}_1,\vec{g}_2,\vec{g}_3,\vec{m} \rangle_{AB} \nonumber \\
& = \hat{\mathcal{S}} \hat{H}^C_{AB} \hat{R}^\dagger \hat{\mathcal{P}}_{A \leftrightarrow C} |\vec{g}_1,\vec{g}_2,\vec{g}_3,\vec{m} \rangle_{BC}, \label{last}
\end{align}
where in the third row we introduce the shorthand notation $ |\vec{g}_1\rangle_{C_1} \otimes |\vec{g}_2\rangle_{C_2} \otimes |\vec{g}_3\rangle_{C_3} \otimes \ket{\vec{m}}_B = |\vec{g}_1, \vec{g}_2, \vec{g}_3, \vec{m} \rangle_{CB}$. In the fifth row we used the fact that $\hat{V}_{C \to A} \hat{U}_{C \to A}$ is unitary and its action as shown in~(\ref{transf_check}). Evidently, $\hat{R}$ stands here for the operator implementing the common rotation~$\hat{M}$ which representation on vectors is~$\cal{M}$. Introducing now a `bra' $\langle\vec{g}_1',\vec{g}_2',\vec{g}_3',\vec{m}' |_{BC}$, and with an analogous procedure, one obtains
\begin{multline}
\langle\vec{g}_1',\vec{g}_2',\vec{g}_3',\vec{m}' | \hat{H}^A_{BC} |\vec{g}_1,\vec{g}_2,\vec{g}_3,\vec{m} \rangle_{BC} = \\
\langle\vec{g}_1',\vec{g}_2',\vec{g}_3',\vec{m}' | \hat{\mathcal{P}}_{A \leftrightarrow C} \hat{R} \hat{H}^C_{AB} \hat{R}^\dagger \hat{\mathcal{P}}_{A \leftrightarrow C} |\vec{g}_1,\vec{g}_2,\vec{g}_3,\vec{m} \rangle_{BC} = \\
\langle\vec{g}_1',\vec{g}_2',\vec{g}_3',\vec{m}' | \hat{\mathcal{P}}_{A \leftrightarrow C} \hat{H}^C_{AB} \hat{\mathcal{P}}_{A \leftrightarrow C} |\vec{g}_1,\vec{g}_2,\vec{g}_3,\vec{m} \rangle_{BC},
\end{multline}
where in the last row we have used the invariance of the Hamiltonian in~(\ref{HamiltonianAB}). This proves~(\ref{HamiltonianBC}).

\bibliography{masterbib}

\end{document}